\documentclass[a4paper,11pt]{article}
\pdfoutput=1 

\usepackage{jheppub} 

\usepackage[T1]{fontenc} 
\usepackage{graphicx}
\usepackage{latexsym}
\usepackage{amssymb,amsmath}

\renewcommand{\a}{\alpha}
\renewcommand{\b}{\beta}

\newcommand{\la}{\lambda}
\newcommand{\La}{\Lambda}
\newcommand{\half}{\frac{1}{2}}

\newcommand{\Om}{\Omega}

\newcommand{\pa}{\partial}
\newcommand{\td}{\textrm{d}}
\newcommand{\nn}{\nonumber\\}

\newcommand{\LM}{\mathcal{L}}
\newcommand{\HM}{\mathcal{H}}

\def\be{\begin{equation}}
\def\ee{\end{equation}}
\def\bea{\begin{eqnarray}}
\def\eea{\end{eqnarray}}
\def\bal{\begin{align}}
\def\eal{\end{align}}
\newcommand{\bit}{\begin{itemize}}
\newcommand{\eit}{\end{itemize}}

\title{\boldmath Hamiltonian analysis of Mimetic gravity with higher derivatives}

\author[a,b,c]{Yunlong Zheng}

\affiliation[a]{Department of Astronomy, School of Physical Sciences, University of Science and Technology of China, Hefei, Anhui 230026, China}
\affiliation[b]{CAS Key Laboratory for Researches in Galaxies and Cosmology, Department of Astronomy, University of Science and Technology of China, Hefei, Anhui 230026, China}
\affiliation[c]{School of Astronomy and Space Science, University of Science and Technology of China, Hefei, Anhui 230026, China}

\emailAdd{zhyunl@ustc.edu.cn}

\abstract{
Two types of mimetic gravity models with higher derivatives of the mimetic field are analyzed in the Hamiltonian formalism. For the first type of mimetic gravity,  the Ricci scalar only couples to the mimetic field and we demonstrate the number of degrees of freedom (DOFs)  is three.
 Then in both Einstein frame and Jordan frame, we perform the Hamiltonian analysis for the extended mimetic gravity with higher derivatives directly coupled to the Ricci scalar. We show that different from previous studies   working at the cosmological perturbation level, where only three propagating  DOFs show up,  this generalized mimetic model, in general, has four DOFs.
To understand this discrepancy, we consider the unitary gauge and find out that the number of DOFs reduces to three. We conclude that the reason why this system looks peculiar is that the Dirac matrix of all secondary constraints becomes singular in the unitary gauge, resulting in extra secondary constraints and thus reducing the number of DOFs. Furthermore, we give a simple example of a dynamic system to illustrate how gauge choice can affect the number of secondary constraints as well as the DOFs when the rank of the Dirac matrix is gauge dependent. 
}

\begin{document}
\maketitle
\flushbottom

\section{Introduction}
\label{sec:intro}
Mimetic scenario  was proposed by Chamseddine and Mukhanov \cite{Chamseddine:2013kea} as a modification of general relativity, where the physics  metric is related to a scalar field and an auxiliary metric via
\be\label{singular}
g_{\mu\nu}=(\tilde{g}^{\a\b}\phi_\a\phi_\b) \tilde{g}_{\mu\nu}~,
\ee
where $\phi_\a\equiv\nabla_\a\phi$ denotes the covariant derivative of the scalar field with respect to spacetime. In this way, the physical metric is invariant under the Weyl rescalings of the auxiliary metric, and the scalar field satisfies the  constraint 
\be\label{mc}
 g^{\mu\nu}\phi_\mu\phi_\nu=1~.
\ee
The  gravitational equations by varying the  Einstein-Hilbert action, which is constructed from the physical metric $g_{\mu\nu}$, contains  an extra scalar mode which can mimic the cold dark matter, hence the theory is dubbed the mimetic dark matter.  

It has been shown that  the number of degrees of freedom remains unchanged under a general invertible disformal transformation \cite{Bekenstein:1992pj},  one may wonder how does the new component arise in the mimetic scenario. The reason is  that mimetic scenario can be viewed as a singular  (non-invertible) limit of general disformal transformation and therefore a new DOF arises in this setup \cite{Deruelle:2014zza,Arroja:2015wpa,Domenech:2015tca}.

Alternatively, the above mimetic constraint can be imposed by employing a Lagrange multiplier in the action \cite{Golovnev:2013jxa}, that is, the action of mimetic gravity takes the following form 
\be
    S=\int\td^4x\left[\half R+\la(g^{\mu\nu}\phi_\mu\phi_\nu-1)\right]+S_m~,
\ee
where $S_m$ is the action for other matter in the universe and we use the most negative signature for the metric. These two formalisms are equivalent, at least classically. We shall take the Lagrange multiplier formalism in this paper, as has been done in most papers of extensions of  mimetic gravity.

The  original mimetic model was generalized in \cite{Chamseddine:2014vna}  by introducing an arbitrary potential. This generalized mimetic model has many applications in cosmology, and can provide us inflation, dark energy, bounce and so on with appropriate choice of the potential $V(\phi)$.
The mimetic constraint can also be applied in various modified gravity models \cite{Nojiri:2014zqa,Leon:2014yua,Astashenok:2015haa,Myrzakulov:2015qaa,
Rabochaya:2015haa,Arroja:2015yvd,Cognola:2016gjy,Nojiri:2016ppu,
Momeni:2014qta,Chamseddine:2018qym,Chamseddine:2018gqh,Alvarez:2018law}, and mimetic gravity has attracted extensive astrophysical and cosmological interests \cite{Saadi:2014jfa,Mirzagholi:2014ifa,Matsumoto:2015wja,Ramazanov:2015pha,Myrzakulov:2015kda,
 Astashenok:2015qzw,Chamseddine:2016uyr,Nojiri:2016vhu,Babichev:2016jzg,Chamseddine:2016ktu, Brahma:2018dwx,
 Sadeghnezhad:2017hmr,Shen:2017rya,Abbassi:2018ywq,Casalino:2018tcd,Vagnozzi:2017ilo,Dutta:2017fjw}.
The Hamiltonian analysis of various mimetic models have also been investigated in Refs.~\cite{Malaeb:2014vua,Chaichian:2014qba,Takahashi:2017pje}, Besides, there are also some other theoretic developments \cite{Hammer:2015pcx,Golovnev:2018icm,Odintsov:2018ggm,Langlois:2018jdg,Firouzjahi:2018xob,Shen:2019nyp,Gorji:2018okn,Paston:2018orc}.
See Ref.  \cite{Sebastiani:2016ras} for a review.

Even being offered a potential, there is no nontrival dynamics for scalar perturbation, i.e. the propagation velocity is zero $c_s=0$. This may rise to  caustic singularities. Besides, the notion of quantum fluctuations is lost as there is no propagating  degree of freedom for the scalar perturbation. Hence, when applied to the early universe, such model fails to produce the primordial perturbations which seeds the formation of large scale structure. To remedy these issues, higher derivative terms $(\Box\phi)^2$ are introduced  in \cite{Chamseddine:2014vna} to promote the scalar degree of freedom to be dynamical with a non-zero sound speed. Although the equation for the scalar perturbation has the wave-like form by choosing appropriate coefficient, the analysis in the action formalism shows that the mimetic scenario with higher derivatives always suffer from ghost instability or gradient instability \cite{Ijjas:2016pad}. Actually, the mimetic model with higher derivative terms can be produced as a certain limit of the projective version of the Horava-Lifshitz gravity and such instability has already been pointed out \cite{Ramazanov:2016xhp}.
It has been shown that simply generalizing the quadratic higher derivative terms to arbitrary function $f(\Box\phi)$ \cite{Firouzjahi:2017txv} or introducing the non-minimal coupling of mimetic field to the Ricci scalar $f(\phi)R$ \cite{Zheng:2017qfs} can not cure this pathology. To find a way out of  the ghost and gradient instabilities, in \cite{Zheng:2017qfs} we  demonstrate that it is possible to circumvent both the ghost and gradient instabilities by introducing the direct couplings of the higher derivatives of the mimetic field to the curvature. Similar couplings are also proposed in  \cite{Hirano:2017zox,Gorji:2017cai}. The extended action in our previous work \cite{Zheng:2017qfs} has the form
\be\label{S0}
S=\int\td^4x\sqrt{-g}\left[\frac{f(\phi,\Box\phi)}{2}R+\la (g^{\mu\nu}\phi_{\mu}\phi_{\nu}-1)-V(\phi)+\a(\Box\phi)^2+\b\phi^{\mu\nu}\phi_{\mu\nu}\right]~.
\ee
From the reduced quadratic action of the perturbations, one scalar and two tensor modes are obtained, and we showed it is indeed possible to avoid both the ghost and gradient instabilities. It seems that we have achieved the goal to construct a healthy model without any instabilities.
However, since the action \eqref{S0} contains the direct coupling between the higher derivative terms of mimetic field and the spacetime curvature, one might be concerned whether the model in general has three DOFs exactly. Besides, the modified dispersion relation \cite{Cai:2009hc} (involving $k^4$ term) of scalar perturbation may imply the existence of extra DOF which do not show up at the perturbation level with cosmological background. The main purpose of this paper is  to identify the number of DOFs for the extended mimetic  model \eqref{S2} which is slightly different from \eqref{S0}. As we shall see,  such kind of theories have four  DOFs for generic field configurations,  and reduce to three after imposing the unitary gauge.

The paper is organized as follows. In the next section, we  perform the full Hamiltonian analysis for the first type of mimetic model, where the
general function of the higher derivative  of the mimetic field  is includes and the Ricci scalar only couples to a function of the mimetic field,
and show the number of DOFs is three, which is consistent with the previous result of the Hamiltonian analysis at perturbation level in \cite{Firouzjahi:2017txv}. In section \ref{sec3}, the full Hamiltonian analysis for the extended mimetic gravity with higher derivatives directly couples to the Ricci scalar is performed in both Einstein frame and Jordan frame, and we find four DOFs in general. To clarify the confusion why only three DOFs show up at the cosmological perturbation level, we also perform the Hamiltonian analysis in the unitary gauge where only three DOFs appear. Finally, we  give a simple example where the rank of the Dirac matrix is gauge dependent in section  \ref{example} followed by conclusion and discussions in section \ref{conclusion}.  Lastly, a special case of mimetic gravity with higher derivative terms is discussed in the Appendix.

\section{Mimetic gravity with higher derivative terms}
We start from the following action of  mimetic theory
\be\label{S1}
	S_1=\int\td^4x\sqrt{-g}\left[\frac{f(\phi)}{2}R+\la (g^{\mu\nu}\phi_{\mu}\phi_{\nu}-1)-V(\phi)+g(\Box\phi)\right]~,
\ee
where $R$ is the Ricci scalar, $\la$ is the Lagrange multiplier enforcing the mimetic constraint \eqref{mc}, $g(\Box\phi)$ is the general higher derivative function and we have considered the non-minimal coupling of the mimetic field to the curvature. This model can be viewed as a generalization of the model in \cite{Firouzjahi:2017txv}, and is  slightly different from the model considered in \cite{Zheng:2017qfs} which includes terms $\phi^{\mu\nu}\phi_{\mu\nu}$.  
Recently, the detecion of the gravitational wave event GW170817 \cite{TheLIGOScientific:2017qsa} has provided strict constraints on the  sound speed of gravitational waves $c_t$, which has to be equal to  the light speed c=1, up to very high accuracy $|c_t^2/c^2-1|\leqslant 5\times10^{-16}$. As one can see from the quadratic action of perturbation in \cite{Zheng:2017qfs}, the inclusion of terms $\phi^{\mu\nu}\phi_{\mu\nu}$ will  change the sound speed of gravitational waves  and  leads to the deviation from the light speed, thus the $\phi^{\mu\nu}\phi_{\mu\nu}$ terms will not be considered in this paper. The main goal of this section is to identify the number of DOFs for the model \eqref{S1}. Introducing a new variable $\varphi=\Box\phi$, one can rewrite the action as
\be
	S_1=\int\td^4x\sqrt{-g}\left[\frac{f(\phi)}{2}R+\la (g^{\mu\nu}\phi_{\mu}\phi_{\nu}-1)-V(\phi)+g(\varphi)+\La(\varphi-\Box\phi)\right]~,
\ee
where  the Lagrange multiplier $\La$ in the last term fixes $\varphi$. To get rid of the appearance of higher derivatives of the mimetic field in the action, we drop the boundary term and simplify the action as 
\be
	S_{1J}=\int\td^4x\sqrt{-g}\left[\frac{f(\phi)}{2}R+\la (g^{\mu\nu}\phi_{\mu}\phi_{\nu}-1)+g^{\mu\nu}\phi_{\mu}\La_{\nu}-V(\phi)+g( \varphi)+\La \varphi\right]~.
\ee
One can switch the action of Jordan frame to the Einstein frame by weyl scaling $g_{\mu\nu}=\Om^2\bar{g}_{\mu\nu}$ where $\Om^2=f(\phi)^{-1}$. The final action in the Einstein frame is
\begin{align}\label{EF1}
	S_{1E}=\int\td^4x\sqrt{-\bar{g}}\bigg[&\frac{\bar{R}}{2}+\frac{3f_\phi^2}{4f^2}\bar{g}^{\mu\nu}\phi_{\mu}\phi_{\nu}
	+\bar{\la} \left(\bar{g}^{\mu\nu}\phi_{\mu}\phi_{\nu}-\frac{1}{f(\phi)}\right)\nn
	&+\frac{1}{f(\phi)}\bar{g}^{\mu\nu}\phi_{\mu}\La_{\nu}+\frac{1}{f(\phi)^2}\left(g(\varphi)+\La\varphi-V(\phi)\right)\bigg]~,
\end{align}
where $\bar{\la}=\la/f(\phi)$. Here, we use bar to distinguish the variables in the Einstein frame from the ones in the Jordan frame. To identify the number of DOFs in this model, we shall perform the full Hamiltonian analysis. Although the Hamiltonian analysis of this model in the case of $f(\phi)=1$ has been studied at the perturbation level \cite{Firouzjahi:2017txv},  there may exist extra DOF not showing up at the perturbation level with cosmological FRW background. Therefore, it is necessary  to perform the Hamiltonian analysis of the general non-perturbation theory.

\subsection{Hamiltonian analysis: Einstein frame}
 In this subsection, we will perform the detailed Hamilonian analysis of the thoery in the Einstein frame. Here we mention that all the bars over the variables, which is used to denote the quantity in the Einstein frame, have been omitted for briefness. Under ADM decomposition, the action \eqref{EF1} becomes
\begin{align}\label{ADMS1}
S_{1E}=\int\td^4x N\sqrt{h}\bigg[&\frac{1}{2}\left(-\mathcal{R}+K^{ij}K_{ij}-K^2\right)+
	\la \left(\frac{(\dot{\phi}-N^i\phi_i)^2}{N^2}-h^{ij}\phi_i\phi_j-\frac{1}{f}\right)\nn
&+\frac{3f_\phi^2}{4f^2} \left(\frac{(\dot{\phi}-N^i\phi_i)^2}{N^2}-h^{ij}\phi_i\phi_j\right) + \frac{1}{f} \left(\frac{(\dot{\La}-N^i\La_i)(\dot{\phi}-N^j\phi_j)}{N^2}-h^{ij}\La_i\phi_j\right)\nn
	&+\frac{1}{f(\phi)^2}\left(g(\varphi)+\La\varphi-V(\phi)\right)\bigg]~,	
\end{align}
where $\mathcal{R}$ denotes the 3-dimensional Ricci scalar and $K_{ij}=(\dot{h}_{ij}-N_{i|j}-N_{j|i})/2N$ is the extrinsic curvature. One can see there are 14 coordinate variables $Q_a=\{N, ~N^i,~h_{ij},~\phi,~\la,~\varphi,~\La\}$. For each coordinate variable $Q_a$, define the conjugate momentum as $\pi_a=\frac{\partial \LM}{\partial \dot{Q}_a}$.  As  the coordinates $~N,~N^i,~\varphi$ and $\la$ have no time derivative in the action, this  leads to six primary constraints
\be\label{primary1}
 \pi_N=0~, ~~\pi_i=0~,~~\Phi_1\equiv\pi_\la=0~,~~\Phi_5\equiv\pi_\varphi=0~.
\ee
Other conjugate momentums are
\begin{align}
\pi^{ij}&=\frac{\sqrt{h}}{2}(K^{ij}-h^{ij} K)~,~~\pi_\La=\frac{\sqrt{h}}{f}\frac{\dot{\phi}-\phi_iN^i}{N}~,\nn
\pi_\phi&=\sqrt{h}\left[\left(\frac{3f_\phi^{~2}}{2f^2}+2\la\right)\frac{\dot{\phi}-\phi_iN^i}{N}+\frac{1}{f}\frac{\dot{\La}-\La_iN^i}{N}\right]~.
\end{align}

Following the standard route , we obtain the total Hamiltonian
\be\label{H1}
H_{1}=\int\td^3x\big[N\mathcal{H}+N^i\mathcal{H}_{i}+v^N\pi_N+v^i\pi_i+v^\varphi\pi_\varphi+v^\la\pi_\la\big]~,
\ee
where
\begin{align}
\mathcal{H}=&\mathcal{H}_g+\mathcal{H}_m=\sqrt{h}\left(\frac{\mathcal{R}}{2}+h^{-1}(2\pi_{ij}\pi^{ij}-\pi^2)\right)+\sqrt{h}\bigg[\frac{f\pi_\phi\pi_\La}{h}-(\la+\frac{3f_\phi^2}{4f^2})\frac{f^2\pi_\La^2}{h}\nn
&+\frac{3f_\phi^2}{4f^2}h^{ij}\phi_i\phi_j+\la(h^{ij}\phi_i\phi_j+\frac{1}{f})+\frac{1}{f}h^{ij}\La_i\phi_j +\frac{1}{f^2}(V(\phi)-g(\varphi)-\La\varphi)\bigg]~,
\end{align}
and
\begin{align}
\mathcal{H}_i=\mathcal{H}_{gi}+\mathcal{H}_{mi}=-2\sqrt{h}\left(\frac{\pi_{~i}^j}{\sqrt{h}}\right)_{|j}+\pi_\phi\phi_i+\pi_\La\La_i+\pi_\varphi\varphi_i+\pi_\la\la_i~.
\end{align}
The conservation of the primary constraints, enables us to determine six corresponding secondary constraints \cite{Cai:2013lqa,Cai:2014upa}. Using Eq. \eqref{H1} together with the primary constraints in \eqref{primary1}, we find
\be
\HM\approx0~,~~\HM_i\approx0~,~~\Phi_2\equiv-\frac{f^2\pi_\La^2}{h}+h^{ij}\phi_i\phi_j+\frac{1}{f(\phi)}\approx 0~,~~\Phi_6\equiv g'(\varphi)+\La\approx0~,
\ee
where the weak equality sign “$\approx$” denotes an identity up to terms that vanish on the constraint surface. By employing the constraint equation $\Phi_6$, one can express $\varphi$ in terms of $\La$. The conservation of constraint $\Phi_6$  determines the Lagrange multiplier $v^\varphi$ and so the chain of constraints for primary constraint $\Phi_5$ terminates here.

Writing the constraints in  smeared form we have
\begin{align}
  \mathrm{H}[N]&=\int\td^3x N(x)\mathcal{H}(x)~,\nn
  \mathrm{D}[N^i]&=\int\td^3x N^i(x)\mathcal{H}_i(x)~.
\end{align}
To recognise that $\mathcal{H}_i$ is indeed the diffeomorphism constraint, we can verify the following Poisson brackets
\begin{align}
 &\{A,\mathrm{D}[N^i]\}=N^iA_{,i}=\mathcal{L}_{\vec{N}}A~,~~
 \{A_i,\mathrm{D}[N^i]\}=A_{i|j}N^j+A_jN^j_{~|i}=\mathcal{L}_{\vec{N}}A_i~,\nn
 &\{\Pi,\mathrm{D}[N^i]\}=(N^i\Pi)_{,i}=\mathcal{L}_{\vec{N}}(\sqrt{h}\frac{\Pi}{\sqrt{h}})~,~~
\{h_{ij},\mathrm{D}[N^i]\}=N_{i|j}+N_{j|i}=\mathcal{L}_{\vec{N}}h_{ij}~,\nn
 & \{\pi^{ij},\mathrm{D}[N^i]\}=N^i_{~|k}\pi^{jk}+N^j_{~|k}\pi^{ik}-\sqrt{h}(\frac{N^k\pi^{ij}}{\sqrt{h}})_{|k}=\mathcal{L}_{\vec{N}}(\sqrt{h}\frac{\pi^{ij}}{\sqrt{h}})~,
\end{align}
where $A$ is a scalar quantity such as $\phi, h^{ij}\phi_i\La_j$ and so on, $A_i$ is a covariant vector quantity such as $\phi_i$, and $\Pi$ is the conjugate momentum quantities such as $\pi_\phi$ or scalar densities with wight $1$ like $\sqrt{h}A $. Here we assume that  $A, A_i, \Pi$ in the above equations only depend on $\phi,\theta, \La, \la, h_{ij} $ and their conjugate momentums (without  $N, N^i$ dependence). Therefore, the Poisson bracket of any constraints $\Phi$  (without  $N, N^i$ dependence) with 
$\mathrm{D}[N^i]$ vanishes after imposing the constraint equation, i.e. $\mathrm{D}[N^i]$ or $\mathcal{H}_i$ is first class. This property greatly simplifies the subsequent process of calculating the secondary constraints.

In addition, the following results are useful to derive the time evolution of variables including constraints 
\begin{align}\label{deltaHN}
	\frac{\delta \mathrm{H}[N]}{\delta\pi_\phi}&=Nf\frac{\pi_\La}{\sqrt{h}}~,
	~~\frac{\delta \mathrm{H}[N]}{\delta\pi_\La}=N\left[\frac{f\pi_\phi}{\sqrt{h}}-2f^2\frac{\pi_\La}{\sqrt{h}}(\la+\frac{3f_\phi^2}{4f^2})\right]~,\nn
	\frac{\delta \mathrm{H}[N]}{\delta \pi^{ij}}&=\frac{2N}{\sqrt{h}}(2\pi_{ij}-h_{ij}\pi)~,~~\frac{\delta \mathrm{H}[N]}{\delta \La}=-\sqrt{h}\left[\left(\frac{N\phi^{|i}}{f}\right)_{|i}+N\frac{\varphi}{f^2}\right]~.
\end{align}
Then the time evolution of constraint $\Phi_2$ reads
\be
	\dot{\Phi}_2\approx\{\Phi_2, H_1\}\approx2N\Phi_3\approx0~,
\ee
where the new constraint
\be
 \Phi_3=\frac{\pi_\La}{\sqrt{h}}\left[f_\phi(h^{ij}\phi_i\phi_j-\frac{3}{2f})-f\phi^{|i}_{~|i}-\varphi(\La)\right]+f h^{ij}\phi_i\left(\frac{\pi_\La}{\sqrt{h}}\right)_{|j}-\frac{2}{\sqrt{h}}(\pi^{ij}\phi_i\phi_j+\frac{\pi}{2f})\approx0~
\ee
can be derived. Futhermore, the next consistency condition generates another new constraint
\be\label{phi4}
 \Phi_4\approx\frac{1}{N}\{\Phi_3,H_1\}=\la\left[\frac{1}{f}(4h^{ij}\phi_i\phi_j+\frac{3}{f})+2\frac{\pa\varphi}{\pa\La}(h^{ij}\phi_i\phi_j+\frac{1}{f})\right] +J_0(\phi,\pi_\phi,\La,\pi_\La,h_{ij},\pi^{ij},D_i)~.
\ee
By requiring the conservation of  the constraint $\Phi_4$, the Lagrange multiplier $v^\la$ is determined  in terms of other variables and so the chain of constraints for the primary constraint $\Phi_1$ terminates here.

Note that $\HM\approx0$ and $\HM_i\approx0$ are expected to correspond to the Hamiltonian and momentum constraints respectively. With some manipulation the following Poisson brackets are found to be the usual ones
\begin{align}
\{\mathrm{D}[\vec{M}],\mathrm{D}[\vec{N}]\}&=\mathrm{D}[\LM_{\vec{M}}\vec{N}]~,\nn
\{\mathrm{D}[\vec{M}],\mathrm{H}[N]\}&=\mathrm{H}[\LM_{\vec{M}}N]~.
\end{align}
We emphasize here that $\HM$ is not first-class, but one can construct a new Hamiltonian constraint $\tilde{\HM}$ \cite{Langlois:2015skt} as a linear combination of $\HM$, $\pi_\la$ and $\pi_\varphi$ such that (up to boundary term)
\be
 \int\td^3x[N\mathcal{H}+v^\varphi\pi_\varphi+v^\la\pi_\la]=\int\td^3xN\tilde{\mathcal{H}}
\ee
where the Lagrange multipliers $v^\varphi$ and $v^\la$ are solved in terms of other variables by requiring all the above consistency conditions. It can be easily seen that $v^\varphi$ and $v^\la$ are linearly dependent on lapse function $N$ or its derivative. Therefore, $N$ is not involved in $\tilde{\HM}$ and the new Hamiltonian constraint $\tilde{\HM}$ is first-class. Besides, it is natural to expect 8 first-class constraints due to the diffeomorphism invariance of the starting theory.
 The time evolution of $\tilde{\HM}$ and  $\HM_i$ do not yield any new constraints and so the chain of constraints for primary constraints $\pi_N$ and $\pi_i$ terminate.

To sum up, the above considerations show that there are 14 constraints:
\bea
  8~\rm{first-class}&:& ~\pi_N~,~\pi_i~,~\tilde{\HM}~,~\HM_i~,\nn
  6~\rm{second-class}&:&~\Phi_1~,~\Phi_2~,~\Phi_3~,~\Phi_4~,~\Phi_5~,~\Phi_6~.
\eea
These constraints reduce the dimension of phace space and thus the number of DOFs for the model \eqref{S1} according to definition by Dirac are
\begin{align}
	\half(28-2\times8-6)=3~,
\end{align}
which is consistent with the Hamiltonian analysis in \cite{Firouzjahi:2017txv,Takahashi:2017pje}.

Besides, there exists a very special case for the general theory \eqref{S1}.  This special case can be found by requiring that $\Phi_4$ doesn't contain $\la$  in \eqref{phi4} in the unitary gauge, i.e.
\be
     \frac{3}{2f}+\frac{\pa \varphi}{\pa \La}=0,
\ee
which gives $f(\phi)=1$ and $g(\Box\phi)=\frac{1}{3}(\Box\phi)^{~2}$ by taking account of the constraint equation $\Phi_6$. As no dependence of $\Phi_4$ on $\la$ in the unitary  gauge will lead to  more secondary constraints than in the general gauge, thus less DOFs show up  in the unitary gauge for this special case. 
More detailed discussion about this special case can be found in the Appendix.

\section{Mimetic gravity with higher derivative terms couples to the curvature}
\label{sec3}
It has been shown \cite{Ijjas:2016pad, Firouzjahi:2017txv, Zheng:2017qfs} that mimetic model like \eqref{S1} suffer from ghost instability or gradient instability. To overcome this difficulty, one can introduce the direct couplings of the higher derivatives of the mimetic field to the curvature of the spacetime. In this section, we shall consider the following extended mimetic model
\be\label{S2}
	S_2=\int\td^4x\sqrt{-g}\left[\frac{f(\phi,\Box\phi)}{2}R+\la (g^{\mu\nu}\phi_{\mu}\phi_{\nu}-1)-V(\phi)+g(\Box\phi)\right]~,
\ee
which is slightly different from the model \eqref{S0} proposed by \cite{Zheng:2017qfs}.
The aim of this section is to identify the number of DOFs for the model \eqref{S2}. Similar to the method in the previous section, one can introduce a new variable $\varphi=\Box\phi$ and rewrite the action as
\be
	S_2=\int\td^4x\sqrt{-g}\left[\frac{f(\phi,\varphi)}{2}R+\la (g^{\mu\nu}\phi_{\mu}\phi_{\nu}-1)-V(\phi)+ g(\varphi)+\La(\varphi-\Box\phi)\right]~.
\ee
To avoid the higher derivative terms of the mimetic field in the action, we drop the boundary term and derive
\be
	S_2=\int\td^4x\sqrt{-g}\left[\frac{f(\phi,\varphi)}{2}R+\la (g^{\mu\nu}\phi_{\mu}\phi_{\nu}-1)+g^{\mu\nu}\phi_{\mu}\La_{\nu}-V(\phi)+ g(\varphi) +\La \varphi\right]~.
\ee
To simplify the calculation one can define a new variable $\chi=f(\phi,\varphi)$ and the inverse function $\varphi=F(\phi,\chi)$, then we acquire
\be\label{JF}
	S_{2J}=\int\td^4x\sqrt{-g}\left[\frac{\chi}{2}R+\la (g^{\mu\nu}\phi_{\mu}\phi_{\nu}-1)+g^{\mu\nu}\phi_{\mu}\La_{\nu}-V(\phi)+g( F(\phi,\chi))+\La F(\phi,\chi)\right]~.
\ee
Switching the action in the Jordan frame to the Einstein frame by weyl scaling $g_{\mu\nu}=\Om^2\bar{g}_{\mu\nu}$ where $\Om^2=\chi^{-1}=\exp{(\frac{2}{\sqrt{6}}\theta)}$, one obtain 
\bal\label{EF}
	S_{2E}=\int\td^4x\sqrt{-\bar{g}}\bigg[&\frac{\bar{R}}{2}+\half\bar{g}^{\mu\nu}\theta_{\mu}\theta_{\nu}+\bar{\la} (\bar{g}^{\mu\nu}\phi_{\mu}\phi_{\nu}-e^{\frac{2\theta}{\sqrt{6}}})+e^{\frac{2\theta}{\sqrt{6}}}\bar{g}^{\mu\nu}\phi_{\mu}\La_{\nu}\nn
	&+e^{\frac{4\theta}{\sqrt{6}}}\left(g(F(\phi,\theta))+\La F(\phi,\theta)-V(\phi)\right)\bigg]~.
\end{align}
We shall first perform the full Hamiltonian analysis in the Einstein frame and then do the similar analysis in the Jordan frame. We will see the results in both frames are consistent with each other.

\subsection{Hamiltonian analysis: Einstein frame}
 The goal of  this subsection is to analyze the number of  DOFs for the model \eqref{S2} in the Jordan frame.  We will first consider general field configurations, and then homogeneous field configurations.
 
\subsubsection{General field configurations}
Here we mention that all the bars over the variables in the Einstein frame have been omitted in this subsubsection for briefness. Then, the action \eqref{EF} becomes

\bal
	S_{2E}=\int\td^4x\sqrt{-g}\bigg[&\frac{R}{2}+\la (g^{\mu\nu}\phi_{\mu}\phi_{\nu}-e^{\frac{2\theta}{\sqrt{6}}})+\half g^{\mu\nu}\theta_{\mu}\theta_{\nu}+e^{\frac{2\theta}{\sqrt{6}}}g^{\mu\nu}\phi_{\mu}\La_{\nu}\nn
	&+e^{\frac{4\theta}{\sqrt{6}}}\big(g(F)+\La F-V(\phi)\big)\bigg]~.
\end{align}
In ADM decomposition, the action takes the form
\begin{align}
S_{2E}=\int\td^4x N\sqrt{h}\bigg[&\frac{1}{2}\left(-\mathcal{R}+K^{ij}K_{ij}-K^2\right)+
	\la \left(\frac{(\dot{\phi}-N^i\phi_i)^2}{N^2}-h^{ij}\phi_i\phi_j-e^{\frac{2\theta}{\sqrt{6}}}\right)\nn
	&+\half \left(\frac{(\dot{\theta}-N^i\theta_i)^2}{N^2}-h^{ij}\theta_i\theta_j\right)  +e^{\frac{2\theta}{\sqrt{6}}} \left(\frac{(\dot{\La}-N^i\La_i)(\dot{\phi}-N^j\phi_j)}{N^2}-h^{ij}\La_i\phi_j\right)\nn
	&+e^{\frac{4\theta}{\sqrt{6}}}\big(g(F)+\La F-V(\phi)\big)\bigg]~.
\end{align}
The coordinate $N$, $N^i$ and $\la$ have no time derivatives in the action, which means we have five primary constraints
\be\label{primary2}
 \pi_N=0~, ~~\pi_i=0~,~~\Phi_1\equiv\pi_\la=0~.
\ee
Other non-vanishing conjugate momentums are defined as
\begin{align}
\pi^{ij}&\equiv\frac{\partial\mathcal{L}}{\partial\dot{h_{ij}}}=\frac{\sqrt{h}}{2}(K^{ij}-h^{ij}K)~,
~~\pi_\theta\equiv\frac{\sqrt{h}}{N}(\dot{\theta}-N^i\theta_i)~,\nn
\pi_\phi&\equiv\frac{\sqrt{h}}{N}[2\la(\dot{\phi}-N^i\phi_i)+e^{\frac{2\theta}{\sqrt{6}}}(\dot{\La}-N^i\La_i)],~~\pi_\La\equiv\frac{\sqrt{h}}{N}e^{\frac{2\theta}{\sqrt{6}}}(\dot{\phi}-N^i\phi_i)~.
\end{align}

After some calculations we obtain the total Hamiltonian
\be\label{H2}
H_{2E}=\int\td^3x[N\mathcal{H}+N^i\mathcal{H}_{i}+v^N\pi_N+v^i\pi_i+v^\varphi\pi_\varphi+v^\la\pi_\la]~,
\ee
where
\begin{align}
\mathcal{H}=\mathcal{H}_g+\mathcal{H}_m=&\sqrt{h}\left(\frac{\mathcal{R}}{2}+h^{-1}(2\pi_{ij}\pi^{ij}-\pi^2)\right)+\sqrt{h}\bigg[(\frac{\pi_\theta^2}{2h}+\half h^{ij}\theta_i\theta_j)+h^{-1}(\pi_\phi\pi_{\La}e^{\frac{-2\theta}{\sqrt{6}}}-\la\pi_{\La}^2e^{\frac{-4\theta}{\sqrt{6}}})\nn
&+\la(h^{ij}\phi_i\phi_j+e^{\frac{2\theta}{\sqrt{6}}})+h^{ij}\La_i\phi_j e^{\frac{2\theta}{\sqrt{6}}}-e^{\frac{4\theta}{\sqrt{6}}}\big(g(F)+\La F-V(\phi)\big)\bigg]~,
\end{align}
and
\begin{align}
\mathcal{H}_i=\mathcal{H}_{gi}+\mathcal{H}_{mi}=-2\sqrt{h}\left(\frac{\pi_i^j}{\sqrt{h}}\right)_{|j}+\pi_\theta\theta_i+\pi_\phi\phi_i+\pi_\varphi\varphi_i+\pi_\la\la_i~.
\end{align}

The time evolution of primary constraints generate the corresponding secondary constrains, which are the Hamiltonian constraint
\be
	\mathcal{H}=\mathcal{H}_g+\mathcal{H}_m\approx0,
\ee
the diffeomorphism constraint
\be
\mathcal{H}_i=\mathcal{H}_{gi}+\mathcal{H}_{mi}\approx0,
\ee
and mimetic constraint
\be\label{MC}
\Phi_2\equiv-\frac{\pi_\La^2}{h}e^{\frac{-4\theta}{\sqrt{6}}}+h^{ij}\phi_i\phi_j+e^{\frac{2\theta}{\sqrt{6}}}\approx0~,
\ee
respectively. Imposing the consistency condition of mimetic constraint yields 
\be\label{Phi2s}
 \{\Phi_2,H_{2E}\}\approx\int\td^3x\left(\frac{\delta \Phi_2}{\delta h_{ij}} \frac{\delta \mathrm{H}_g[N]}{\delta\pi^{ij}} +\frac{\delta \Phi_2}{\delta\theta} \frac{\delta \mathrm{H}_m[N]}{\delta\pi_\theta} +\frac{\delta \Phi_2}{\delta\phi} \frac{\delta \mathrm{H}_m[N]}{\delta\pi_\phi}-\frac{\delta \Phi_2}{\delta\pi_\La} \frac{\delta \mathrm{H}_m[N]}{\delta\La} \right)\approx 0~,
\ee
here we have used the property of diffeomorphism constraint $\{\Phi_2,\mathrm{D}[N^i]\}=N^i\Phi_{2,i}\approx 0$. To obtain the new secondary constraint,
one need to first compute the following useful functional derivatives of $\Phi_2$ 
\begin{align}\label{Dphi2}
  \frac{\delta\Phi_2[g]}{\delta h_{ij}}&=g\sqrt{h}(\frac{\pi_{\La}^{~2}}{h}e^{\frac{-4\theta}{\sqrt{6}}} h^{ij}-\phi^i\phi^j)~,
  ~~\frac{\delta\Phi_2[g]}{\delta\phi}=-\sqrt{h}(2g\phi^i)_{|i}~, \nn
  \frac{\delta\Phi_2[g]}{\delta\theta}&=\frac{2}{\sqrt{6}}g\sqrt{h}(2\frac{\pi_{\La}^{~2}}{h}e^{\frac{-4\theta}{\sqrt{6}}}+e^{\frac{2\theta}{\sqrt{6}}})~,
  ~~\frac{\delta\Phi_2[g]}{\delta\pi_\La}=-\frac{2g\pi_\La}{\sqrt{h}}e^{\frac{-4\theta}{\sqrt{6}}}~,
\end{align}
and other useful functional derivatives of  Hamiltonian
\begin{align}\label{Dham}
 \frac{\delta \mathrm{H}_g[N]}{\delta\pi^{ij}}&=\frac{4N}{\sqrt{h}}(\pi_{ij}-\frac{\pi}{2}h_{ij})~,~~
 \frac{\delta \mathrm{H}_m[N]}{\delta\pi_\theta}=\frac{N\pi_\theta}{\sqrt{h}}~,\nn
 \frac{\delta \mathrm{H}_m[N]}{\delta\pi_\phi}&=\frac{N\pi_\La}{\sqrt{h}}
 e^{-\frac{2\theta}{\sqrt{6}}}~,
 ~~\frac{\delta \mathrm{H}_m[N]}{\delta\pi_\La}=\frac{N}{\sqrt{h}}(\pi_\phi e^{-\frac{2\theta}{\sqrt{6}}} -2\la \pi_\La e^{-\frac{4\theta}{\sqrt{6}}})~,\nn
 \frac{\delta \mathrm{H}_m[N]}{\delta\la}&=N\sqrt{h}\Phi_2~,
 ~~\frac{\delta \mathrm{H}_m[N]}{\delta\La}=-\sqrt{h}[(N\phi^i e^{\frac{2\theta}{\sqrt{6}}})_{|i}+N F e^{\frac{4\theta}{\sqrt{6}}}]~.
\end{align}
Plugging the above formulae into \eqref{Phi2s} we have
\be
	\{\Phi_2,H_{2E}\}=2N\Phi_3\approx0~,
\ee
where the new constraint reads
\begin{align}\label{phi3}
  \Phi_3(\phi,\theta,\pi_\theta,\pi_\La,h_{ij},\pi^{ij})=&-\frac{2}{\sqrt{h}}(\frac{\pi}{2}e^{\frac{2\theta}{\sqrt{6}}}+\pi^{ij}\phi_i\phi_j)-\frac{\pi_\La}{\sqrt{h}}F 
  + \frac{\pi_\theta}{\sqrt{6h}}(\frac{2\pi_\La^2}{h}e^{-\frac{4\theta}{\sqrt{6}}}+e^{\frac{2\theta}{\sqrt{6}}})\nn
   &-e^{-\frac{2\theta}{\sqrt{6}}}\left[\frac{\pi_\La}{\sqrt{h}}\phi^i_{~|i}-\phi^i\left(\frac{\pi_\La}{\sqrt{h}}\right)_{|i} +\frac{4\theta}{\sqrt{6}}\phi^i\theta_i\frac{\pi_\La}{\sqrt{h}} \right]~.
\end{align}
With the constraint equations $\HM,~\Phi_2,~\Phi_3$, one can express $\pi_\phi,~\pi_\La,~\pi_\theta$ in terms of other variables and thus we can eliminate the dependence on $\pi_\phi,~\pi_\La,~\pi_\theta$ in the later calculation.

We now  compute the following useful Poisson bracket
\be\label{PB23}
 \{\Phi_2(y),\Phi_3(z)\}=\int\td^3x\left(\frac{\delta \Phi_2(y)}{\delta h_{ij}(x)} \frac{\delta \Phi_3(z)}{\delta\pi^{ij}(x)}+ \frac{\delta \Phi_2(y)}{\delta \theta(x)} \frac{\delta \Phi_3(z)}{\delta\pi_\theta(x)}\right)~.
\ee
The relevant functional derivatives of $\Phi_3$ are given by
\begin{align}\label{Dphi3}
    \frac{\delta \Phi_3(y)}{\delta\pi^{ij}(x)}&=-\frac{2}{\sqrt{h}}(\frac{h_{ij}}{2}e^{\frac{2\theta}{\sqrt{6}}}+\phi_i\phi_j) ~\delta^3(y-x)~,\nn
	\frac{\delta \Phi_3(y)}{\delta \pi_\theta(x)}&= \frac{1}{\sqrt{6h}}(\frac{2\pi_\La^2}{h}e^{-\frac{4\theta}{\sqrt{6}}}+e^{\frac{2\theta}{\sqrt{6}}})~\delta^3(y-x)~.
\end{align}
Plugging (\ref{Dphi2}) and \eqref{Dphi3} into \eqref{PB23}, one has
\be
\{\Phi_2(y),\Phi_3(z)\}\approx~\frac{4}{3\sqrt{h}}(\nabla\phi)^4~\delta^3(y-z)~,
\ee
where the constraint equation $\Phi_2\approx 0$ is used.  The time evolution of $\Phi_3$ leads to another new constraint
\be
	\{\Phi_3,H_{2E}\}=N\Phi_4\approx 0~,
\ee
where the new constraint is
\be
 \Phi_4=-\frac{4}{3}\la (\nabla\phi)^4+J(\phi,\theta,\La,h_{ij},\pi^{ij},D_i)~.
\ee
Here the explicit expression of $J$ function is tediously long and not important for our purpose. The key point is that the direct calculation shows $\Phi_4$ does not depend on $N$.
Because of the dependence of $\Phi_4$ on $\la$, the time evolution of $\Phi_4$ involves Lagrange multiplier $v^\la$, thus the chain of constraints for primary constraint $\Phi_1$ terminates here. Similar to the previous section, the time evolution of $\mathcal{H},\mathcal{H}_i$ are automatically satisfied and yield nothing.

The above considerations show that five primary constraints $\{\Phi_1, \pi_N, \pi_i\}$ yield  seven secondary constraints $\{\Phi_2,\Phi_3,\Phi_4,\mathcal{H},\mathcal{H}_i\}$, 
therefore we have  the standard 8 first-class constraints and 4 second-class constraints  in all. According to the usual definition of DOFs for constraint systems, the number of DOFS in the theory \eqref{S2} is \be
14-8-\half\times4=4~.
\ee

However, as
\be
\{\Phi_1,\Phi_4\}=\frac{4}{3}(\nabla\phi)^4\delta^3(x-y)~,
\ee
one has $\{\Phi_1,\Phi_4\}=0$ if the mimetic scalar field is homogeneous $\nabla\phi=0$. Meanwhile, $\Phi_1$ commutes with all the other constraints. We can see that  the Dirac matrix of the 4 secondary constraints $\{\Phi_1, \Phi_2, \Phi_3, \Phi_4\}$ will become singular (non-invertible) and the rank will reduce by two if  we take the homogeneous field background $\nabla\phi=0$ (which is related to the coordinate choice).  In the case of homogeneous field configuration, we have to redone the analysis because the conservation of $\Phi_4$ will yield further constraints rather than fix $v^{\lambda}$.
This actually implies that the number of DOFs will become three in the unitary gauge,  as will be shown in the subsubsection below.  

\subsubsection{Unitary gauge}
If we consider our model $S_{2E}$  in the special unitary gauge $\phi=t$  from the beginning, i.e. the effective field theory (EFT) $S_{2E}^{(u)}=S_{2E}-\int\td^4x~u(\phi-t)$, and then do the similar Hamiltonian  analysis as above, we can obtain the new total Hamiltonian
\be\label{Hu}
	H_{2E}^{(u)}=H_{2E}+\int\td^3x~u(\phi-t)~,
\ee
which is just the former Hamiltonian plus one additional term imposing the unitary gauge condition. The primary constraints now are given by
\be
 \pi_N\approx0~,~~\pi_i\approx 0~,~~\tilde{\Phi}_1\equiv\pi_\la\approx0~,~~\tilde{\Phi}_7\equiv\phi-t\approx0~.
\ee
Here we use tilde to distinguish the constraints in the unitary gauge from the ones in arbitrary gauge. The time evolution of those constraints generate the following secondary constraints
\be
 \mathcal{H}\approx0,~~\mathcal{H}_i\approx0, ~~\tilde{\Phi}_2\equiv=-\frac{\pi_\La}{\sqrt{h}}+e^{\frac{3\theta}{\sqrt{6}}}\approx0, ~~\tilde{\Phi}_8\equiv N-e^{\frac{-\theta}{\sqrt{6}}}\approx0.
\ee
where these expressions have been simplified by employing the constraints equation. Requiring $\tilde{\Phi}_8$ to be time independent gives
\be
   v^N+\frac{\pi_\theta}{\sqrt{h}}e^{\frac{-\theta}{\sqrt{6}}}\approx 0~,
\ee
which determines the Lagrange multiplier $v^N$ and so the chain of constraints for $\tilde{\Phi}_7$ terminate here.

The time evolution of mimetic constraint $\tilde{\Phi}_2$ gives us a new constraint
\be
  \tilde{\Phi}_3(\phi,\theta,\pi_\theta,h_{ij},\pi^{ij})=-\pi-\sqrt{h} F e^{\frac{\theta}{\sqrt{6}}}
  + \frac{3}{\sqrt{6}}\pi_\theta\approx0~.
\ee
Through a direct calculation, we find out that $\{\tilde{\Phi}_2,\tilde{\Phi}_3\}\approx 0$ just as expected.  The time evolution of $\tilde{\Phi}_3$ generates a new constraint
\be
	\tilde{\Phi}_4(\phi,\theta,\La,h_{ij},\pi^{ij},D_i)\approx\{\tilde{\Phi}_3,H_{2E}^{(u)}\}\approx0~,
\ee
which is independent of $\la$. With the constraint equations   $\HM,~\tilde{\Phi}_2,~\tilde{\Phi}_3$ , one can  express $\pi_\phi,~\pi_\La,~\pi_\theta$ in terms of other variables and thus eliminate the dependence on $\pi_\phi,~\pi_\La,~\pi_\theta$ in the later calculation. Through a direct calculation,  one find out that the Poisson bracket of $\tilde{\Phi}_2$ and $\tilde{\Phi}_4$  weakly vanish. Another new  constraint generated by the next consistency relation is derived to be
\be
	\tilde{\Phi}_5(\phi,\theta,\La,h_{ij},\pi^{ij},D_i)\approx\{\tilde{\Phi}_4,H_{2E}^{(u)}\}~,
\ee
 which also has no dependence on $\la$.
The exact expression of $\tilde{\Phi}_4$ and $\tilde{\Phi}_5$ is complicated, but fortunately for our purpose we only care which variables they depend on. As $\la$ is not involved in $\tilde{\Phi}_5$,  the  time evolution of  $\tilde{\Phi}_5$ yield another constraint $\tilde{\Phi}_6$ depending on $\la$. Therefore, the time evolution of $\tilde{\Phi}_6$ involves the Lagrangian multiplier $v^\la$,   and the chain of constraints for $\tilde{\Phi}_1=\pi_\la\approx0$ terminates here.

Furthermore, as we have taken the unitary gauge which breaks the first-class property of energy constraint, the time evolution of $\HM$ gives $u=0$ while the time evolution of  $\mathcal{H}_i$
are still automatically satisfied. Therefore  the chain of constraints for $\pi_N$ and $\pi_i$  terminate here.

Above considerations show that six primary constraints yield  nine secondary constraints in total, therefore the system admits 16 constraints which are
\bea
  6~\rm{first-class}&:& ~\pi_i~,~\HM_i~,\nn
  10~\rm{second-class}&:&~\pi_N~,~\HM~,~\tilde{\Phi}_1~,~\tilde{\Phi}_2~,~\tilde{\Phi}_3~,~\tilde{\Phi}_4~,~\tilde{\Phi}_5~,~\tilde{\Phi}_6~,~\tilde{\Phi}_7~,~\tilde{\Phi}_8~.
\eea
According to the usual counting of DOFs for constraint systems, the number of independent  DOFs for the setup \eqref{Hu} is $14-6-\half\times10=3$~.
We emphasize here that the number of DOFs  is indeed different between the general case and the homogeneous field configurations. Normally it is  supposed that gauge choice should not affect the physics and the number of DOFs. What is special in our theory is that the associated Dirac matrix happens to  be singular for the unitary gauge, resulting in further constraints ($\tilde{\Phi}_5, \tilde{\Phi}_6$) besides the usual  unitary gauge fixing conditions ($\tilde{\Phi}_7, \tilde{\Phi}_8$).

As one can always choose the gauge invariant quantities to describe the perturbations of the system, the cosmological perturbation theory should be the same between the general case and the homogeneous field configurations. This leads to the conclusion that we can only see 3 degrees of freedom (1 scalar and 2 tensor modes) in the perturbation theory of our model $S_2$, and  the other one scalar degree of freedom don't appear in the cosmological background. This is  consistent with our previous paper \cite{Zheng:2017qfs}, which works in the Lagrangian formalism and only consider the perturbation theory at the  isotropic and homogeneous  background. 

\subsection{Hamiltonian analysis: Jordan frame}
The aim of this subsection is to obtain the number of  DOFs of the model \eqref{S2} in the Jordan frame and compare it with the result in the Einstein frame. 

\subsubsection{General field configurations}
Starting with the action \eqref{JF} in the Jordan frame,  one can rewrite it in the ADM formalism
\begin{align}
S_{2J}=\int\td^4x N\sqrt{h}\bigg[&\frac{\chi}{2}\left(-\mathcal{R}+K^{ij}K_{ij}-K^2\right)-\frac{\dot{\chi}-N^i\chi_i}{N}K+\frac{h^{ij}\chi_i N_{,j}}{N}\nn
	&+\la \left(\frac{(\dot{\phi}-N^i\phi_i)^2}{N^2}-h^{ij}\phi_i\phi_j-1\right)
	+\left(\frac{(\dot{\La}-N^i\La_i)(\dot{\phi}-N^j\phi_j)}{N^2}-h^{ij}\La_i\phi_j\right)\nn
	&+ g(F(\phi,\chi))+\La F(\phi,\chi)-V(\phi)\bigg]~.
\end{align}
As the action does not include time derivatives of $N$, $N^i$, and $\la$, one immediately has the primary constraints
\be\label{primary2}
	\pi_N\approx0~,~\pi_i\approx0~,~\Psi_1\equiv\pi_\la\approx0~.
\ee
Here we use $\Psi_A$ to denote the constraints in Jordan frame. Other conjugate momentums are
\begin{align}
\pi^{ij}&=\frac{\partial \mathcal{L}}{\partial \dot{h_{ij}}}
=\frac{\sqrt{h}}{2}\left[\chi(K^{ij}-h^{ij}K)-\frac{\dot{\chi}-N^k\chi_k}{N}~h^{ij}\right]~,\nn
\pi_\phi&=\sqrt{h}\left(2\la~\frac{\dot{\phi}-N^i\phi_i}{N}+\frac{\dot{\La}-N^i\La_i}{N}\right)~,
\pi_\La=\sqrt{h}\frac{\dot{\phi}-N^i\phi_i}{N}, \pi_\chi=-\sqrt{h}K~.
\end{align}

The total Hamiltonian is then given by
\be
H_{2J}=\int\td^3x\left[N\mathcal{H}+N^i\mathcal{H}+v^N\pi_N+v^i\pi_i+v^\la\pi_\la\right]~,
\ee
where
\begin{align}
\mathcal{H}=&\frac{1}{\sqrt{h}}\left[\pi_\phi\pi_\La-\la\pi_\La^2+\frac{\chi}{3}\pi_\chi^2-\frac{2}{3}\pi\pi_\chi+\frac{2}{\chi}(\pi_{ij}\pi^{ij}-\frac{1}{3}\pi^2)\right]\nn
&+\sqrt{h}\left[\frac{\chi}{2}\mathcal{R}+\chi^{|i}_{~|i}+\la(h^{ij}\phi_i\phi_j+1)+h^{ij}\phi_i\La_j-g(F)-\La F+V(\phi)\right]
\end{align}
and
\be
\mathcal{H}_i=\pi_\phi\phi_i+\pi_\La\La_i+\pi_\chi\chi_i+\pi_\la\la_i-2\sqrt{h}\left(\frac{\pi_i^{~j}}{\sqrt{h}}\right)_{|j}~.
\ee
With the primary constraints \eqref{primary2},  we find the corresponding secondary constraints to be the Hamiltonian constraint, diffeomorphism constraint and mimetic constraint
\be
  \mathcal{H}\approx0~,~\mathcal{H}_i\approx0~,~\Psi_2=-\frac{\pi_\La^2}{h}+h^{ij}\phi_i\phi_j+1\approx0~.
\ee

Again, one can write the constraints in smeared form as before. We will frequently use the property in the subsequent calculations that the Poisson bracket of any constraint (without  $N, N^i$ dependence) with $\mathrm{D}[N^i]$ vanishes, i.e. $\{\Psi,\mathrm{D}[N^i]\}=\mathcal{L}_{\vec{N}}\Psi\approx0$.
The following functional derivatives will be useful for the subsequent calculations
\begin{align}
 &\frac{\delta\mathrm{H}[N]}{\delta\pi^{ij}}=\frac{2N}{\sqrt{h}}\left[\frac{2}{\chi}(\pi_{ij}-\frac{\pi}{3}h_{ij})-\frac{\pi_\chi}{3}h_{ij}\right]~,
  ~\frac{\delta\mathrm{H}[N]}{\delta\pi_\phi}=\frac{N\pi_\La}{\sqrt{h}}, ~\frac{\delta\mathrm{H}[N]}{\delta\pi_\La}=\frac{N}{\sqrt{h}}(\pi_\phi-2\la\pi_\La)~,\nn
 &\frac{\delta\mathrm{H}[N]}{\delta\pi_\chi}=\frac{2N}{3\sqrt{h}}(\chi\pi_\chi-\pi)~, ~\frac{\delta\mathrm{H}[N]}{\delta\La}=-\sqrt{h}[(N\phi^{|i})_{|i}+NF]~, \nn
 &\frac{\delta\mathrm{H}[N]}{\delta\chi}=\frac{N}{\sqrt{h}}\left[\frac{\pi_\chi^2}{3}-\frac{2}{\chi^2}(\pi_{ij}\pi^{ij}-\frac{\pi^2}{3})\right]+N\sqrt{h}\left[\frac{\mathcal{R}}{2}-(g_{,F}+\La)F_{,\chi}\right]+\sqrt{h}N^{|i}_{~|i}~.
\end{align}

The time evolution of mimetic constraint is
\be
 \{\Psi_2,H_{2J}\}=\int\td^3x\left(\frac{\delta \Psi_2}{\delta h_{ij}} \frac{\delta \mathrm{H}[N]}{\delta\pi^{ij}} +\frac{\delta \Psi_2}{\delta\phi} \frac{\delta \mathrm{H}[N]}{\delta\pi_\phi}-\frac{\delta \Psi_2}{\delta\pi_\La} \frac{\delta \mathrm{H}[N]}{\delta\La} \right)=2N\Psi_3\approx 0~,
\ee
where we have used $\{\Psi_2,\mathrm{D}[N^i]\}\approx0$. The explicit expression of the new constraint is
\be
	\Psi_3=-\frac{\pi_\La}{\sqrt{h}}(\phi^i_{|i}+F)+h^{ij}\phi_i(\frac{\pi_\La}{\sqrt{h}})_{|j}-\frac{\pi_\La^2}{h}\frac{\pi_\chi}{\sqrt{h}}-\frac{2}{\chi\sqrt{h}}(\pi^{ij}-\frac{\pi}{3}h^{ij})\phi_i\phi_j+ \frac{\pi_\chi}{3\sqrt{h}}h^{ij}\phi_i\phi_j~.
\ee
With the constraint equations $\HM,~\Psi_2,~\Psi_3$ ,  one can express $\pi_\phi,~\pi_\La,~\pi_\theta$ in terms of other variables and thus we can  eliminate the dependence on $\pi_\phi,~\pi_\La,~\pi_\theta$ in the later calculation.

The time evolution of $\Psi_3$ leads to another constraint
\be
 \{\Psi_3,H_{2J}\}\approx N\Psi_4\approx0~.
\ee
By using the result of the following Poisson bracket
\be
	\{\Psi_2(y),\Psi_3(z)\}=\frac{4}{3\sqrt{h}\chi}(\nabla\phi)^4\delta^3(y-z)~,
\ee
the new constraint is obtained to be
\be
	\Psi_4=-\frac{4\la}{3\chi}(\nabla\phi)^4+J_2(\phi,\chi,\La,h_{ij},\pi^{ij},D_i)~.
\ee
Here the explicit expression of $J_2$ is tedious and not important for us. The key point is that through direct calculation we find  all the terms involving $N$ cancel exactly, i.e. $\Psi_4$ does not depend on $N$. Requiring this constraint to be time independent, determines the Lagrangian multiplier $v^\la$ in terms of phase space variables and the chain of constraints for primary constraint $\Psi_1=\pi_\la\approx0$ terminates. Besides, the time evolution of $\mathcal{H}_i$ are automatically satisfied and yield no extra constraint.

Therefore the system admits 12 constraints $\Psi_A=\{\Psi_1,\Psi_2,\Psi_3,\Psi_4,\pi_N,\mathcal{H},\pi_i,\mathcal{H}_i\}$,
 of which 8 are  first class and 4 are second class. Thus, the number of independent physical degrees of freedom in the model \eqref{S2} is $14-8-\half\times4=4$ which is consistent with the analysis in the Einstein frame. Similar to the Einstein frame discussed above, one can see that if the mimetic field is homogeneous $\nabla\phi=0$,  the Dirac matrix will become singular and  the rank will reduce. We shall work out the Hamiltonian analysis under the unitary gauge to check the change of DOFs.

\subsubsection{Unitary gauge}
Consider the action $S_{2J}$ in the special unitary gauge $\phi=t$, and
then one obtain the new total Hamiltonian
\be\label{Hu}
	H_{2J}^{(u)}=H_{2J}+\int\td^3x ~u(\phi-t)~,
\ee
which is  the former Hamiltonian plus one additional term imposing the unitary gauge condition. The primary constraints now are given by
\be
 \pi_N\approx0~,~~\pi_i\approx 0~,~~\tilde{\Psi}_1\equiv\pi_\la\approx0~,~~\tilde{\Psi}_7\equiv\phi-t\approx0~.
\ee
The time evolution of those constraints generate the following new constraints
\be
 \mathcal{H}\approx0~,~~\mathcal{H}_i\approx0~, ~~\tilde{\Psi}_2\equiv-\frac{\pi_\La^2}{h}+1\approx0~, ~~\tilde{\Psi}_8\equiv N-1\approx0~,
\ee
where these expressions have been simplified by using the constraints equation. Requiring $\tilde{\Psi}_8$ to be time independent gives
\be
   v^N+\frac{\pi_\theta}{\sqrt{h}}e^{\frac{-\theta}{\sqrt{6}}}\approx 0~,
\ee
which determines the Lagrange multiplier $v^N$ and so the chain of constraints for unitary gauge $\tilde{\Psi}_7$ terminates here.

The time evolution of mimetic constraint $\tilde{\Psi}_2$ generates a new constraint
\be
  \tilde{\Psi}_3(\phi,\theta,\pi_\theta,h_{ij},\pi^{ij})=-(\pi_\La F+\pi_\chi)\approx0~.
\ee
One finds $\{\tilde{\Psi}_2,\tilde{\Psi}_3\}\approx 0$ just as expected. The time evolution of $\tilde{\Psi}_3$ yields a new constraint
\be
	\tilde{\Psi}_4(\phi,\chi,\La, h_{ij},\pi^{ij},D_i)\approx\{\tilde{\Psi}_3,H_{2J}^{(u)}\}\approx0~.
\ee
One can verify that  the Poisson bracket of $\tilde{\Psi}_2$ and $\tilde{\Psi}_4$ vanishes just as the case in the Einstein frame. The time evolution of this new constraint $\tilde{\Psi}_4$ also gives another constraint
\be
	\tilde{\Psi}_5(\phi,\chi,\La,h_{ij},\pi^{ij},D_i)=\{\tilde{\Psi}_4,H_{2J}^{(u)}\}~,
\ee
which has no dependence on $\la$. Although the exact expression of $\tilde{\Psi}_4$ and $\tilde{\Psi}_5$ is complicated, the key point is $\tilde{\Psi}_4$ and $\Psi_5$ do not include $\la$ and the  time evolution of  $\tilde{\Psi}_5$ yield another constraint $\tilde{\Psi}_6$ involving $\la$. Therefore, $\dot{\tilde{\Psi}}_6\approx0$ involves the Lagrangian multiplier $v^\la$ and so the chain of constraints for $\tilde{\Psi}_1=\pi_\la\approx0$ terminates here.

Further more, as we have set the unitary gauge which satisfy $\{\phi-t,\mathrm{H}[N]\}\neq 0$, the time evolution of $\mathcal{H}$ determines $u=0$ and so the chain of constraints for $\pi_N\approx0$ terminates here. The first class property of spatial diffeomorphism is unspoiled in the unitary gauge, and the time evolution of $\HM_i$ is automatically preserved, so the  chain of constraints for $\pi_i$ terminates here.

Above considerations show that six primary constraints yield  nine secondary constraints in total, therefore the system admits 16 constraints which are
\bea
  6~\rm{first-class}&:& ~\pi_i~,~\HM_i~,\nn
  10~\rm{second-class}&:&~\pi_N~,~\HM~,~\tilde{\Psi}_1~,~\tilde{\Psi}_2~,~\tilde{\Psi}_3~,~\tilde{\Psi}_4~,~\tilde{\Psi}_5~,~\tilde{\Psi}_6~,~\tilde{\Psi}_7~,~\tilde{\Psi}_8~.
\eea
Therefore, the number of independent  DOFs for the Hamiltonian \eqref{Hu} is $14-6-\half\times10=3$, which is different from the result of general field configurations without gauge fixing. 

\subsection{Homogeneous field configurations and degrees of freedom}
In the above two subsections, we have shown that the number of DOFs is the same between the Einstein frame and the  Jordan frame,  thus our result is independent of the frame as expected. More explicitly, We find that the number of DOFs   for the general theory \eqref{S2} is 4, and reduce to 3 under the unitary gauge. The extra DOF seems to be automatically eliminated  by the homogeneous field configurations. 

Actually, this situation has been encountered and studied in other modified gravity models like Horava-Lifshitz gravity, Cuscuton models and U-degenerate theories \cite{Blas:2009yd,Gomes:2017tzd,DeFelice:2018ewo}.
In \cite{Gomes:2017tzd}, the Hamiltonian analysis for the Cuscuton shows that, for general field configurations, there are three DOFs which reduce to two after taking the unitary gauge. They argued that both cases have to be considered as two different dynamical systems as the homogeneous limit is singular and discontinuous from the general case. 
In \cite{DeFelice:2018ewo}, U-degenerate theories was proposed to represent Higher-Order Scalar-Tensor theories (HOST) which appear degenerate when restricted to the unitary gauge but are not degenerate in the generic gauge. 
It was argued that for degenerate HOST models the additional non-propagating ghost DOF in a generic gauge can be removed by imposing the regularity condition, while this boundary condition is already imposed implicitly  in the unitary gauge.  When the gradien of the scalar field is timelike and with appropriate boundary conditions, the systems propagate a single scalar DOF, while the extra shadowy mode is nondynamical.
This  means that, the theories are immune from Ostrogradski instabilities within these conditions and therefore worth exploring phenomenologically. We shall come back to  this topic in the future.

\section{Gauge dependence of the rank of the Dirac matrix: a simple example}
\label{example}
To better understand the relation between gauge choice and the number of DOFs, here we consider a simple constrained system with countable DOFs in which the rank of the Dirac matrix is gauge dependent.  The idea is to construct a Hamiltonian  system with one first-class $f_1$ (gauge system) and two second-class constraints $f_2$ and $f_3$, and the rank of the associated Dirac matrix is related to the gauge choice, i.e. 
\bea
  f_1  \rm{~is ~first-class~}&:& ~ \{f_1,f_2\}\approx0,\quad \{f_1,f_3\}\approx0,\nn
  f_2 \rm{~and~} f_3   \rm{~are~second-class~}&:& ~ D_{23}= \{f_2,f_3\}\not\approx0~,\nn
  \rm{gauge~dependence ~of~ the ~rank}&:&~\{f_1, D_{23}\}=\{f_1, \{f_2,f_3\}\}\not\approx0~,
\eea
where the sign $\not\approx$ denotes an inequality up to terms that vanish  on the constraint surface.

To realize such an idea, we can first assume the following total Hamiltonian of the system
\be\label{exa}
 H_T=H(q^a,p_a)+u f_1+v f_2,\quad a=1, ...,N
\ee
where $f_1$,$f_2$ are two primary constraints and $u,v$ are the corresponding Lagrange multiplier.  As  $f_1$ is supposed to be first-class, we require 
\be
	\{f_1,f_2\}\approx0,~\{f_1,H\}\approx0.
\ee
The evolution of the  constraint $f_1$ is automatically satified and yields no new constraint, while the time evolution of the constraint $f_2$ generates a new constraint $f_3=\{f_2,H\}$. Requiring $f_3$ to be time independent gives
\be
     \{f_3,H\}+v\{f_3,f_2\}=0.
\ee
As $f_2,f_3$ are second-class, $D_{23}=\{f_3,f_2\}$ is generally not vanishing on the constraint surface and the above equation  determines the Lagrange multipier $v$. Therefore, the number of DOFs is $(2N-2\times1-2)/2=N-2$.  

We further assume that the rank of  the associated Dirac matrix is  gauge dependent, i.e. $\{f_1, D_{23}\}\not\approx0$.  When we take the following special gauge for the system
\be\label{gf}
D_{23}=0,
\ee
the new total Hamiltonian becomes
\be\label{HTgf}
   H_T^{gauge}=H_T+w D_{23}=H(q^a,p_a)+u f_1+v f_2+w D_{23}
\ee 
where the Lagrange multiplier $w$ enforces the gauge fixing \eqref{gf}. Thus, we have 3 primary constraints
\be
 f_1\approx0,~f_2\approx0,~D_{23}\approx0.
\ee
The time evolutions of $f_1$ determines $w=0$,   the time evolutions of $D_{23}$ involves Lagrange multiplier $u$,  while the time evolutions of $f_2$ yield the secondary constraint  $f_3\approx0$. The consistency relation of $f_3$ generate a new constraint
\be
  f_4\equiv\{f_3,H\}\approx0
\ee 
after using the gauge condition $D_{23}\approx0$. 

Thus in the gauge fixing \eqref{gf},  we have at least 5 constraints $\{f_1, f_2, f_3, f_4, D_{12}\}$ while only 4 constraints exist in arbitrary gauge.  Such a simple example is a good demonstration that for some systems  there exist a special gauge which leads to extra secondary constraints  and less DOFs. 

\section{Conclusion and discussions}
\label{conclusion}
Recently, there is an increasing investigation in exploring the instability issue \cite{Ijjas:2016pad} of mimetic model with higher derivative terms. In the previous work \cite{Zheng:2017qfs} we pointed out that it is possible to overcome this pathology by introducing the direct couplings of the higher
derivatives of the mimetic  field to the Ricci scalar of the spacetime. Although it seems that the model in \cite{Zheng:2017qfs} have one scalar mode and two tensor modes by analyzing the quadratic actions of perturbations, the inclusion of $k^4$ for the modified dispersion relation of scalar perturbation may imply the existence of extra DOF which do not show up at the cosmological  perturbation level. 
In this paper we first confirmed that the mimetic models  \eqref{S1} 
has 3 DOFs which is consistent with the results of the Hamiltonian analysis in \cite{Firouzjahi:2017txv} and \cite{Takahashi:2017pje}. Besides, we find a special case where  $f(\phi)=1$ and $g(\Box\phi)=\alpha(\Box\phi)^{2}$. Then we perform a detailed Hamiltonian constraint analysis for the extended mimetic model \eqref{S2} (which is slightly different from the model considered in \cite{Zheng:2017qfs}) in both Einstein frame and Jordan frame. The results are consistent with each other in both frames. Such kind of theories in generic scalar field has 4  DOFs while only 3 propagating DOFs appear  at the cosmological perturbation level \cite{Zheng:2017qfs}. To clarify the discrepancy, we reanalyze  the model after fixing the homogenous scaclar field gauge. Interestingly, the number of DOFs reduces to 3. Therefore, we  concluded that the number of DOFs for the model \eqref{S2}  is 4 in general, and the extra DOF is automatically eliminated by the homogeneous field configurations. We also mention that this interesting situation has been encountered and investigated in a series of modified gravity models.

This also give us a clue why there exist some spatially covariant gravity models including 3DOFs, such as the XG3 theory \cite{Gao:2014soa,Gao:2018znj}, even broader than the DHOST theory \cite{Langlois:2015cwa, BenAchour:2016fzp}.  A reasonable explanation is that extra DOFs may come out from the XG3 theory when recovering the spacetime diffeomorphism. 

Furthermore,   we should point out the appearance of higher power of $w$ and $k$ than two in the dispersion relation  (such as in the case of the XG3 theory  and Horava gravity \cite{Horava:2009uw}) may suggests the existence of extra non-propagating DOF. The relation between the modification of  dispersion relation and the existence for extra DOF  deserves detailed investigations in the future.

Note added: after the submission of this paper to arXiv, we note that a new paper \cite{Ganz:2019vre} appeared. The authors in \cite{Ganz:2019vre} discussed mimetic gravity theories with direct couplings between the curvature and higher derivatives of the scalar field only up to the quintic order. They first acquired effective field theories with three DOFs, and found that there are in general four or more DOFs for a generic scalar field.  Besides, they studied linear scalar perturbations around flat space for inhomogeneous background scalar field configurations, and they demonstrated that there are indeed two scalar modes.

\acknowledgments
I would like to thank  Mingzhe Li,  Liuyuan Shen, Haoming Rao, Dehao Zhao, Tan Liu for useful discussions, and I am  also grateful to  Xian Gao, Karim Noui and Yifu Cai  for helpful corresppondences. This work is supported in part by  NSFC under Grant No. 11847239, 11961131007, 11422543, 11653002, 11421303, and by the Fundamental Research Funds for the Central Universities.

\appendix
\section{a special case}
Here we consider the case where  $f(\phi)=1$ and $g(\Box\phi)=\alpha(\Box\phi)^{2}$. The action $S_1$ reduces to
\be\label{A1}
   S_1=\int\td^4x\sqrt{-g}\left[\frac{1}{2}R+\la (g^{\mu\nu}\phi_{\mu}\phi_{\nu}-1)-V(\phi)+\alpha(\Box\phi)^{2}\right]~.
\ee
In the special case $\alpha=\frac{1}{3}$, one has the Poisson bracket $\{\Psi_2,\Psi_3\}\propto (\nabla\phi) ^2$, and $\Psi_4=\la (\nabla\phi) ^2+J_0$. As $\Psi_4$ involves $\la$, the time evolution will fix the Lagrangian multiplier $v^\la$ and the chain of constraints for $\pi_\la$ terminates here. The number of DOFs is three according to previous analysis. But if we set the unitary gauge from the beginning, the evolution of $\Psi_4$ will generate two secondary constraints $\Psi_5$ and $\Psi_6$. This will reduce the number of DOFs to  2 which means there are only two tensor modes at the corresponding perturbation theory.

The quadratic action for scalar perturbation in the model \eqref{A1} is \cite{Ijjas:2016pad}
\be
   S^{(2)}_\zeta=\int\td^4x a^3 \left[-\frac{1-3\a}{\a}\dot{\zeta}^2+\frac{(\nabla\zeta)^2}{a^2}\right]~,
\ee
where the  time derivative term happens to vanish in the special case $\alpha=\frac{1}{3}$. Then, the equation of motion gives $\zeta=0$.
Therefore, only two tensor perturbation modes contribute to the DOFs. However,  the background equation in this special case becomes \cite{Ijjas:2016pad}
\be \label{BG3}
  0=V(t),
\ee
which will be  self-consistent only if the model have no potential term. But if  the potential is vanishing, the background equation \eqref{BG3} will be automatically  satisfied and gives us nothing about the evolution of the scale factor at all.

\end{document}